\documentclass[5p,times,twocolumn]{elsarticle}

\usepackage{graphicx}
\usepackage{amsmath,amssymb,bm}
\usepackage{booktabs}
\usepackage{multirow}
\usepackage{array}
\usepackage{siunitx}
\usepackage{xcolor}
\usepackage{url}
\usepackage{hyperref}
\usepackage{microtype}
\usepackage{placeins}
\usepackage{enumitem}
\hypersetup{colorlinks=true,linkcolor=blue,citecolor=blue,urlcolor=blue}
\journal{Fusion Engineering and Design}

\newcommand{\pelm}{P_{\mathrm{ELM}}}
\newcommand{\vturb}{v_{\mathrm{turb}}}

\newcommand{\mhz}{\,\mathrm{MHz}}

\begin{document}

\begin{frontmatter}

\title{A packet-level digital hardware twin for commissioning megahertz diagnostic edge AI and plasma control system integration in tokamaks}

\author[uw]{Semin~Joung\corref{cor1}}
\ead{semin.joung@wisc.edu}
\author[slac]{A.~Dave}
\author[slac]{L.~Scomparin}
\author[uw]{F.~Khabanov}
\author[uw]{Z.~Yan}
\author[uw]{B.~Geiger}
\author[uw]{G.~McKee}
\author[ga]{M.~Fantuzzi}
\author[ga]{B.~Penaflor}
\author[ga]{D.~Piglowski}
\author[ga]{S.~P.~Smith}
\author[slac]{R.~Coffee}
\author[uw]{D.~R.~Smith\corref{cor1}}
\ead{david.smith@wisc.edu}
\cortext[cor1]{Corresponding author}
\address[uw]{Department of Nuclear Engineering and Engineering Physics, University of Wisconsin--Madison, Madison, WI, USA}
\address[slac]{SLAC National Accelerator Laboratory, Menlo Park, CA, USA}
\address[ga]{DIII-D National Fusion Facility, General Atomics, San Diego, CA, USA}

\begin{abstract}
High-bandwidth plasma diagnostics increasingly provide inputs to machine learning and signal-processing algorithms intended for real-time tokamak control, but the complete path from diagnostic sampling to control-system handoff is difficult to commission because of their sampling rates. We develop a packet-level digital hardware twin for the megahertz diagnostic edge-AI architecture. The simulator represents a 64-channel, $1\mhz$ beam emission spectroscopy (BES) diagnostic embedded in a 96-channel dual-carrier acquisition system, two 48-channel streams with SPAD0 sample counters, 10~GbE Hardware UDP transport, packet loss and network jitter, FPGA parsing and dual-carrier alignment, causal preprocessing and edge inference, a compact Ethernet result packet, receiver-side shared state, and a 1~kHz PCS-like control cycle. Binary UDP payloads and PCAP files are generated rather than emulating transport only at the array level. With a baseline of 20 samples per packet, each carrier generates 50,000 packets~s$^{-1}$ and 100~MB~s$^{-1}$ of user payload. A 110~ms reference run produces 11,000 HUDP packets; an intentionally dropped 20-sample packet is detected by the sample-counter continuity logic and invalidates the two overlapping 128-sample inference windows without silent interpolation. For valid windows, the configured engineering latency model gives a median last-input-to-shared-memory latency of 91.6~$\mu$s and a 99th percentile of 108.1~$\mu$s. A separate operating-system loopback test sends binary FPGA-result datagrams through a UDP receiver into POSIX shared memory and preserves packet sequence and CRC for 20/20 packets. Interactive GUI interfaces expose timing, packetization, network faults, inference thresholds, and control-state inspection. The framework provides a reproducible environment for testing diagnostic-to-accelerator interfaces and fail-safe behavior for deployment on fusion devices.
\end{abstract}


\end{frontmatter}

\section{Introduction}
\label{sec:intro}

Real-time plasma control increasingly depends on diagnostic information that is both high dimensional and sampled at frequencies far above the native cycle rates of conventional tokamak control systems. This trend is driven by two related objectives. First, advanced control requires direct access to fluctuation, profile, and imaging measurements that were historically available only after a discharge. Second, machine learning (ML) observers can reduce these measurements to compact state estimates that are sufficiently fast for feedback, provided that acquisition, transport, inference, and control-system handoff are engineered as a complete deterministic path rather than as independent components \cite{kim2026rtmonitoring,margo2024pcs,abbate2023infrastructure}.
 
DIII-D provides an important example of this transition. The recently reported real-time plasma monitoring framework integrates high-bandwidth diagnostics with the Plasma Control System (PCS) through the SHIELD architecture and dedicated processing nodes \cite{kim2026rtmonitoring}. That work demonstrated that full-resolution real-time BES can agree with the corresponding off-line data and emphasized the role of local FPGA processing for sub-millisecond quantities.

The remaining engineering challenge addressed here occurs earlier in the deployment cycle. A new acquisition or accelerator path must often be designed before the final hardware, firmware, receiver process, and control-room integration are simultaneously available. The risk is distributed across many interfaces: facility clock and trigger interpretation, ADC quantization, packet payload layout, sequence and sample-counter continuity, dual-link deskew, packet loss and reordering, FPGA window validity, preprocessing causality, result-packet versioning, receiver-side stale-state detection, and controller fail-safe behavior. Testing only the ML model or only the network bandwidth does not expose failures that emerge from the interaction of these layers.

This project is developing an edge-AI path intended to stream megahertz diagnostic data directly to a programmable accelerator and forward only compact control-relevant outputs to PCS. The current DIII-D-oriented engineering baseline considered in this work uses two D-TACQ ACQ2206-class carriers, each populated for 48 fast channels and paired with 10~GbE Hardware UDP (HUDP) transport, to provide a 96-channel acquisition capacity. Sixty-four channels are assigned to BES in the present simulator. D-TACQ documents the ACQ2206 as a six-site carrier with optional MGT483/MGT483-10G communications and HUDP streaming, and recommends a sample counter in SPAD0 so that a receiver can recover data position and identify missing samples \cite{dtacq_acq2206,dtacq_hudp,dtacq_udpx}. The simulator developed here turns those hardware contracts into an executable representation that continues through an FPGA-like receiver and into a PCS-facing software boundary.

We use the term \emph{digital hardware twin} in a restricted sense. The model is an executable representation of the data, timing, packet, accelerator, and control-system interfaces of the planned path. The synthetic plasma response and the ELM-probability observer are included only to exercise downstream interfaces and controller state transitions. This distinction is important because the principal contribution is an engineering commissioning environment.

The main contributions are as follows. First, we implement binary dual-carrier HUDP packetization, including ADC codes, SPAD0 counters, Ethernet/IPv4/UDP encapsulation, PCAP export, and deterministic reconstruction of a synchronized 96-channel stream. Second, we propagate packet-level faults through FPGA window validity, causal preprocessing, compact inference output, receiver state, and PCS-like control logic. Third, we provide a versioned binary result packet and an operating-system loopback mode using UDP sockets and POSIX shared memory so that process boundaries can be exercised before the final DIII-D receiver implementation is available. Fourth, the simulator is coupled to interactive frontends for parameter sweeps, waveform inspection, packet decoding, and fault injection. Finally, we quantify the baseline packet rates, fault detection, and configured latency distribution and identify the remaining interfaces that should be replaced by measured hardware behavior before deployment.

The remainder of this paper is organized as follows. Section~\ref{sec:system} defines the DIII-D reference architecture and the boundary between established DIII-D systems and the proposed project path. Section~\ref{sec:simulator} describes the simulator, packet contract, timing model, FPGA-side processing, and PCS-facing interface. Section~\ref{sec:verification} presents packet-level verification, fault injection, latency modeling, and process-level loopback tests. Section~\ref{sec:gui} describes the interactive commissioning interface. Section~\ref{sec:discussion} discusses the relation to the deployed DIII-D framework, limitations, and the steps required to transition from a software twin to hardware-backed commissioning. Conclusions are given in Section~\ref{sec:conclusion}.

\section{DIII-D reference architecture and simulator boundary}
\label{sec:system}

\subsection{Existing DIII-D real-time diagnostic infrastructure}

The present work is aligned with the existing DIII-D real-time monitoring framework. The deployed SHIELD system reported in Ref.~\cite{kim2026rtmonitoring} provides modular real-time acquisition, shared-memory transport, specialized processing nodes, and PCS integration for several diagnostics. SHIELD transfers diagnostic data into shared-memory ring buffers and allows the PCS to read at its own cycle rate. The deployed system therefore establishes both the relevance and feasibility of high-bandwidth diagnostic control on DIII-D.

We explore a complementary diagnostic-to-accelerator path whose critical transport is external HUDP rather than the AICC/SHIELD acquisition path. The two architectures should therefore be compared at the level of function rather than assumed to be identical at the hardware level. Both seek to preserve a high-rate diagnostic stream outside the central PCS until the data have been reduced to a compact control-relevant state. The simulator focuses on the proposed D-TACQ-to-FPGA path and on the handoff from the edge accelerator to a PCS-facing receiver process.

\subsection{Modeled hardware topology}

\begin{figure*}[t]
    \centering
    \includegraphics[width=0.98\textwidth]{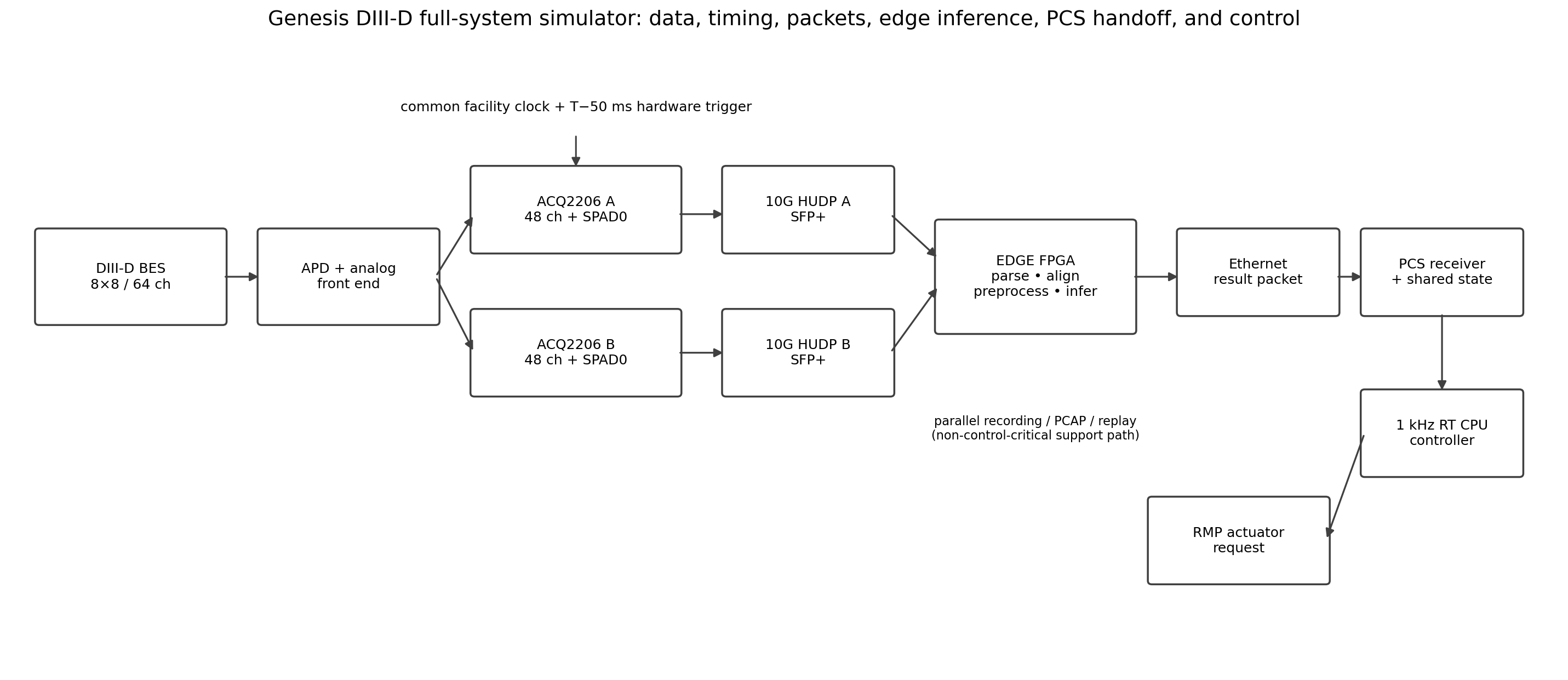}
    \caption{DIII-D-oriented full-system simulator. The modeled control-critical path begins with 64-channel BES, uses two 48-channel acquisition carriers with SPAD0 sample counters and 10~GbE HUDP transport, reconstructs a synchronized stream at an edge FPGA, and forwards a compact result packet through a PCS-facing receiver and shared state to a 1~kHz virtual controller. Recording and replay are parallel support paths. The architecture represents the proposed path.}
    \label{fig:architecture}
\end{figure*}

Figure~\ref{fig:architecture} shows the complete system represented by the simulator. Synthetic BES voltages pass through a simplified APD/electronics stage, 16-bit ADC quantization, two 48-channel acquisition carriers, HUDP packetization, and two independent network links. The FPGA model receives both links, validates and aligns sample counters, constructs the active 64-channel BES tensor, applies causal preprocessing, and evaluates edge outputs. The output is encoded into a compact Ethernet datagram and consumed by a PCS receiver that writes shared state for a 1~kHz control task. Recording, PCAP capture, and replay are treated as parallel support functions and are not placed on the hard real-time control path.

Table~\ref{tab:fidelity} separates the three fidelity levels in the model. Publicly documented or project-grounded quantities are kept distinct from engineering choices that can be swept and from explicit simulator-only placeholders. This separation is used throughout the software and is also exposed in the GUI.

\begin{table*}[t]
\centering
\caption{Fidelity levels used in the digital hardware twin simulator.}
\label{tab:fidelity}
\begin{tabular}{p{0.18\textwidth} p{0.34\textwidth} p{0.40\textwidth}}
\toprule
Level & Examples & Treatment in the simulator \\
\midrule
Project-grounded / vendor-documented & 64 active BES channels; 96-channel planned capacity; two 48-channel carriers; 1~MSPS effective rate; HUDP transport; SPAD0 sample counter; FPGA-to-receiver-to-shared-state concept & Fixed by the baseline configuration but retained as named parameters so that the implementation can be updated with the final hardware configuration. \\
Configurable engineering assumptions & samples per packet, MTU, network jitter, deskew depth, causal filter band, inference latency, PCS cycle, stale timeout & User-adjustable values used for commissioning sweeps and sensitivity tests; no claim that they are measured DIII-D latencies. \\
Simulator-only placeholders & synthetic ELM probability model, virtual plasma/RMP response, PCS Packet v0 binary layout & Isolated behind explicit interfaces so that trained models, measured actuator behavior, and the final DIII-D PCS packet/shared-memory schema can replace them without changing the surrounding data path. \\
\bottomrule
\end{tabular}
\end{table*}

\section{Full-system simulator}
\label{sec:simulator}

\subsection{Synthetic BES and ADC representation}

The source stage is a physics-motivated synthetic BES signal generator rather than a first-principles turbulence simulation. It uses a fixed representative $8\times8$ DIII-D-like viewing geometry and constructs propagating spatially coherent fluctuations, broadband fluctuations, precursor-like amplitude changes, event-centered bursts, channel gain variation, and common/local detector noise. The resulting 64-channel signal is AC only and sampled at $f_s=1\mhz$. This is consistent with the role of BES as a high-rate fluctuation measurement and with the 64-channel, 128-sample windows previously used for ELM-prediction studies \cite{mckee1999bes,joung2024elm,joung2026tokamakagnostic}.

The analog voltage $V_i[n]$ of channel $i$ is quantized to a signed 16-bit code,
\begin{equation}
q_i[n] = \mathrm{clip}\!\left(\mathrm{round}\!\left[\frac{V_i[n]}{V_{\mathrm{FS}}}\left(2^{15}-1\right)\right],-2^{15},2^{15}-1\right),
\label{eq:adc}
\end{equation}
where $q_i[n]$ is the stored ADC code for sample index $n$ and $V_{\mathrm{FS}}$ is the positive full-scale voltage. The 64 active BES channels are embedded in a 96-channel hardware vector; unused channels are present in the packetized carrier data but are not passed to the BES inference block.

\subsection{Dual-carrier HUDP packetization and timing}

The 96 hardware channels are divided into two carriers, A and B, each containing 48 simultaneous 16-bit values and one 32-bit SPAD0 word per sample. For $N_{\mathrm{SPAD}}$ SPAD words and $N_{\mathrm{SPP}}$ samples per packet, the user-payload length is
\begin{equation}
L_{\mathrm{UDP}}=N_{\mathrm{SPP}}\left(48\times2+4N_{\mathrm{SPAD}}\right) \quad \mathrm{bytes}.
\label{eq:payload}
\end{equation}
Here, $N_{\mathrm{SPP}}$ is the number of samples per packet and $N_{\mathrm{SPAD}}$ is the number of 32-bit sample metadata words. The baseline uses $N_{\mathrm{SPP}}=20$ and $N_{\mathrm{SPAD}}=1$, giving a 2000-byte HUDP user payload. At 1~MSPS, the resulting packet rate is 50,000 packets~s$^{-1}$ per carrier and the user-payload throughput is 100~MB~s$^{-1}$ per carrier. The simulator checks the payload against the configured MTU before a run begins.

Each SPAD0 value is the sample counter. The scientific time base is reconstructed from the trigger epoch rather than packet arrival time,
\begin{equation}
t_{\mathrm{DIII-D}}(C)=t_{\mathrm{trig}}+\frac{C-C_{\mathrm{trig}}}{f_s}-\tau_{\mathrm{ADC/FIR}},
\label{eq:time}
\end{equation}
where $C$ is the sample counter, $C_{\mathrm{trig}}$ is the counter associated with the facility trigger, $t_{\mathrm{trig}}$ is the DIII-D trigger epoch represented in the simulator as the $T-50$~ms reference, and $\tau_{\mathrm{ADC/FIR}}$ is a configurable acquisition/filter group delay. Packet arrival times are stored separately and affect only readiness, deskew, latency, and stale-state decisions. This distinction prevents network jitter from corrupting the reconstructed diagnostic time axis.

The simulator serializes each HUDP payload as little-endian ADC codes and SPAD words, constructs Ethernet/IPv4/UDP frames, and writes PCAP files that can be inspected with standard packet-analysis software. The current implementation uses standard UDP encapsulation rather than a custom network shim, allowing the same packet decoder to be exercised either in fast discrete-event mode or through a real local UDP socket.

\subsection{Network impairments and dual-carrier reconstruction}

Each carrier is assigned an independent latency distribution. Packet drop, additional delay, counter jump, random loss, duplication, and reordering probabilities are configurable. At the receiver, samples are indexed by SPAD0 rather than by packet arrival order. A 96-channel sample is considered valid only when both carriers provide the same counter value. The alignment stage therefore implements the conceptual condition
\begin{equation}
\begin{aligned}
C_A[n]=C_B[n] &\Rightarrow \mathrm{valid},\\
C_A[n]\neq C_B[n] &\Rightarrow \mathrm{deskew/invalid}.
\end{aligned}
\label{eq:align}
\end{equation}
where $C_A$ and $C_B$ denote the recovered sample counters of carriers A and B. The simulator never fills a missing packet by interpolation in the real-time path. Missing samples instead propagate as validity information to every 128-sample inference window that overlaps the missing interval.

\begin{figure*}[t]
\centering
\includegraphics[width=0.96\textwidth]{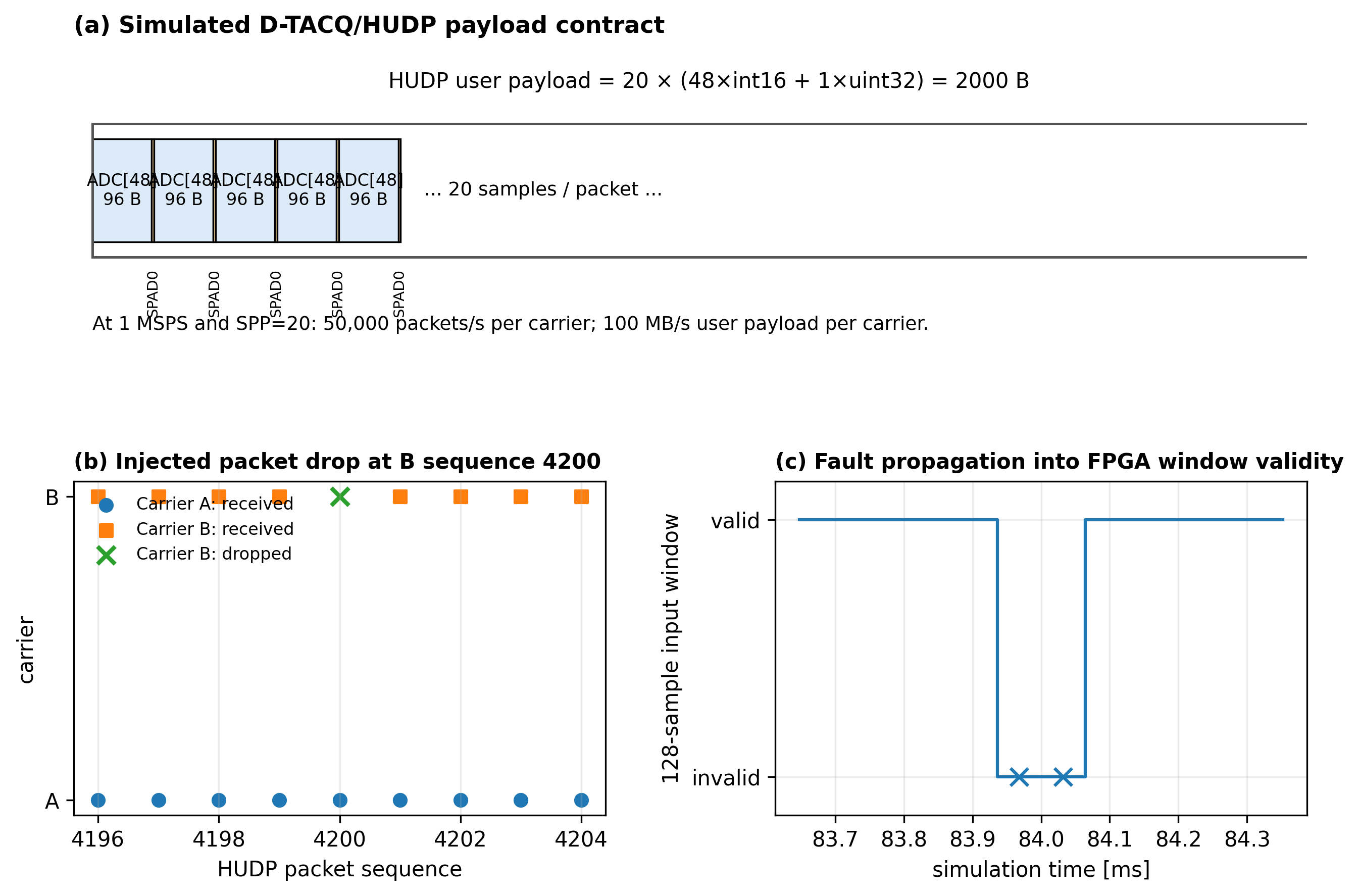}
\caption{Packet contract and fault propagation in the baseline simulation. (a) One HUDP user payload contains 20 samples, each composed of 48 16-bit ADC values and a 32-bit SPAD0 sample counter, for 2000 bytes per packet. (b) Carrier-B packet sequence 4200 is intentionally dropped. (c) The missing 20-sample interval invalidates the two 128-sample, 64-stride FPGA windows that overlap the missing counters. All results in this figure are generated by the simulator rather than measured on DIII-D hardware.}
\label{fig:packetfault}
\end{figure*}

Figure~\ref{fig:packetfault} summarizes the baseline binary contract and one deterministic fault injection. A single dropped packet on carrier B removes counters 84000--84019. Because the FPGA input uses 128-sample windows with a stride of 64, exactly two overlapping inference windows are marked invalid. The example illustrates the intended commissioning behavior: packet loss becomes an explicit health state instead of an unobserved change to the ML input.

\subsection{Causal preprocessing and edge inference}

After carrier alignment, the 64 active BES channels are converted back to voltage units and processed with a causal band-pass filter. The baseline passband is 8--180~kHz. For each 128-sample window, the simulator evaluates channel RMS, a spatial coherence proxy, the RMS of temporal differences, a fourth-moment statistic, and a cross-correlation time-delay velocity proxy. For two vertically separated channels with separation $\Delta z$, the CCTD-like estimate is
\begin{equation}
\vturb = \frac{\Delta z}{\tau^*},
\label{eq:cctd}
\end{equation}
where $\tau^*$ is the lag that maximizes the discrete cross-correlation within a bounded lag search. Several adjacent-channel pairs are evaluated and combined by a median to reduce sensitivity to a single pair.

A lightweight logistic surrogate converts the synthetic features to $\pelm$. It is trained only on labels derived from the synthetic event schedule and is therefore not a scientific ELM predictor. Its function is to reproduce the interface behavior of a compact model: a scalar probability and turbulence-velocity-like state become available at each valid window. The trained DIII-D/KSTAR edge models developed in previous work \cite{joung2024elm,joung2026tokamakagnostic,gill2024regime} can later replace this surrogate through the same inference interface.

\subsection{Versioned FPGA-to-PCS packet and shared state}

The final DIII-D FPGA-to-PCS binary contract was not available when this simulator was developed. Rather than inventing a format and treating it as facility truth, the simulator isolates a versioned 64-byte \emph{PCS Packet v0}. The packet contains a magic/version field, health flags, shot identifier, sequence number, sample counter, reconstructed DIII-D timestamp, $\pelm$, $\vturb$, RMS, coherence, a confidence proxy, and a CRC32. Invalid FPGA windows produce packets with invalid health flags and sentinel model values.

The packet is transported over a configurable standard-Ethernet model to a receiver process, decoded, CRC checked, and written to a single shared state. A generation counter is incremented at each successful write. This design follows the project-level interface assumption of a compact FPGA output reaching a receiver application that publishes data for the PCS, while keeping the detailed packet layout replaceable.

The PCS-like task executes at 1~kHz. At cycle time $t_k$, the state is considered usable when the most recent write is CRC-valid and younger than a configurable timeout $T_{\mathrm{stale}}$,
\begin{equation}
\mathcal{V}(t_k)=\left[\mathrm{CRC}=1\right]\wedge\left[t_k-t_{\mathrm{write}}\leq T_{\mathrm{stale}}\right].
\label{eq:validstate}
\end{equation}
Here, $\mathcal{V}$ is the control-state validity flag and $t_{\mathrm{write}}$ is the most recent receiver write time. If the state becomes stale, the baseline controller holds the previous actuator request; an idle fail-safe mode is also available. The control logic and virtual RMP response are simulator placeholders intended only to exercise state transitions.

\subsection{Discrete-event and process-level execution modes}

Two execution modes are provided. The first is a fast discrete-event mode in which packet serialization, network delay, FPGA timing, receiver delay, and PCS reads are represented as timestamped events. This mode can generate thousands of packets quickly and is used for most parameter scans. The second mode uses actual operating-system UDP sockets and Python POSIX shared memory. In this mode, FPGA-result datagrams cross a process boundary, the receiver decodes and verifies the packet, and a separate reader observes the shared-memory generation counter. The process-level mode does not reproduce hard-real-time scheduling, but it verifies binary compatibility and software boundaries that are absent in a purely array-based simulation.

\section{Verification and commissioning tests}
\label{sec:verification}

\subsection{Baseline packet rate and continuity}

The reference simulation duration is 110~ms. At 1~MSPS with 20 samples per HUDP packet, each carrier produces 5500 packets, giving 11,000 packets in total. The two links therefore each carry 100~MB~s$^{-1}$ of HUDP user payload before Ethernet/IP/UDP overhead. This rate is well below the 10~GbE capability documented for the ACQ2206/MGT483-10G HUDP path \cite{dtacq_hudp,dtacq_udpx}; the limiting commissioning issue in the baseline is therefore packet-rate handling, counter continuity, and deterministic processing rather than raw link bandwidth.

Binary encode/decode unit tests verify HUDP sample reconstruction and the 64-byte result-packet CRC. PCAP export is performed for both the diagnostic-to-FPGA traffic and FPGA-to-PCS result traffic. These files provide a bridge between the simulator and future packet captures from the physical hardware: a measured capture can be introduced at the packet-decoder boundary without changing the downstream processing code.

\subsection{Injected packet drop}

The baseline fault removes one carrier-B packet at sequence 4200, corresponding to counters 84000--84019. Of the 1717 edge-inference windows generated in the reference run, 1715 remain valid and two are invalid, corresponding to 0.116\% invalid windows. Their window-center times are 83.968 and 84.032~ms. No samples are silently synthesized to repair the missing packet.

This short fault does not force a PCS hold in the 1~kHz baseline because subsequent valid edge packets update the shared state before the next relevant control read. This outcome is useful because it shows that diagnostic packet validity and controller stale-state behavior are related but not identical. A packet error must be propagated accurately, while the controller should respond according to the age and health of the latest available state rather than immediately equating any diagnostic loss with an actuator fault. Longer dropouts can be generated from the same interface to test the hold-last or idle fail-safe policies.

\subsection{Configured latency budget}

\begin{figure*}[t]
\centering
\includegraphics[width=0.96\textwidth]{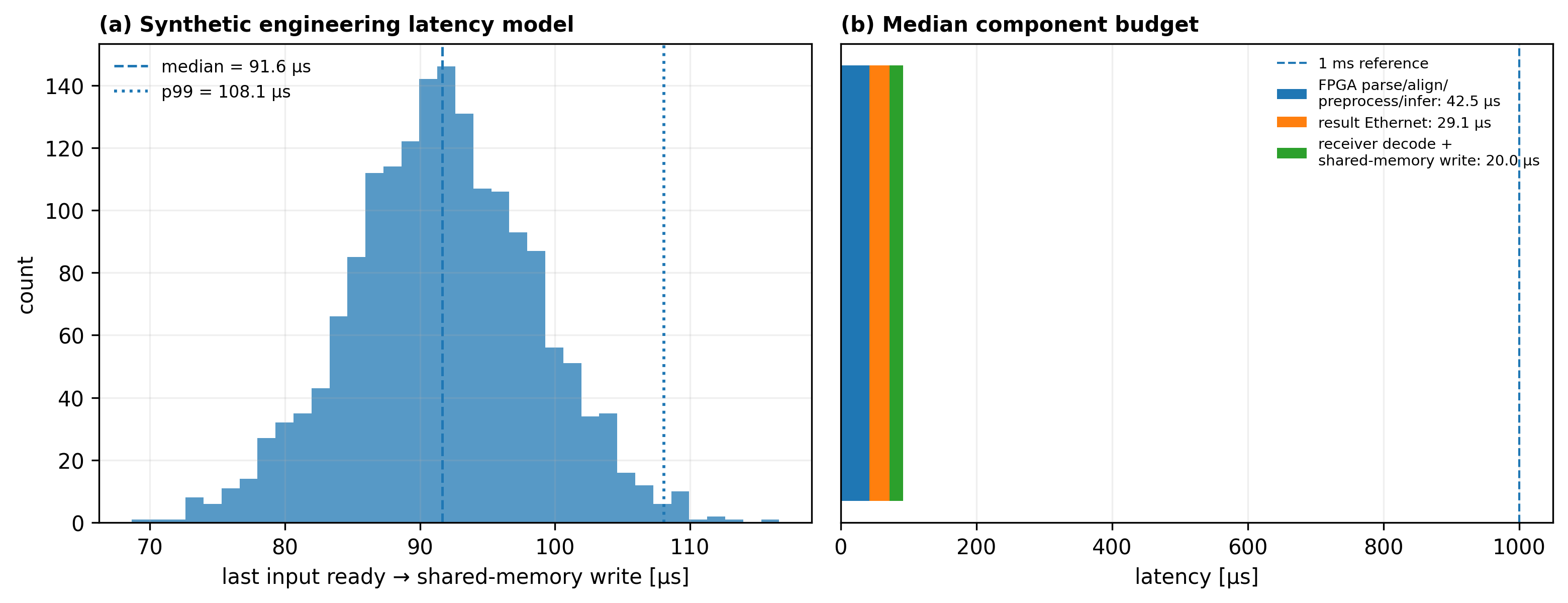}
\caption{Configured engineering latency for valid windows. (a) Distribution of last-input-ready to shared-memory-write latency for the 110~ms baseline run. (b) Median component budget. The 1~ms line is a reference requirement, not a measured hardware comparison. All component delays in this figure are configurable simulator assumptions.}
\label{fig:latency}
\end{figure*}

Figure~\ref{fig:latency} shows the configured latency distribution from the instant the last input sample of a valid window is available at the FPGA receiver to completion of the shared-memory write. The baseline FPGA-side budget is 42.5~$\mu$s, composed of fixed parse, alignment, preprocessing, inference, and result-packet construction delays. The result Ethernet delay is drawn from a normal distribution with a 28~$\mu$s mean and 7~$\mu$s standard deviation, and the receiver decode plus shared-memory write contributes 20~$\mu$s. Across the 1715 valid windows, the resulting median is 91.6~$\mu$s, the 95th percentile is 103.2~$\mu$s, the 99th percentile is 108.1~$\mu$s, and the maximum in the reference realization is 116.6~$\mu$s.

These values are not measured DIII-D or FPGA latencies. They define an engineering budget against which measured component delays can later be substituted. The observation window itself is also kept separate: the 128-sample context corresponds to 128~$\mu$s at 1~MSPS, whereas Fig.~\ref{fig:latency} begins when the last sample of that window becomes available. This distinction prevents the acquisition context length from being confused with compute and transport latency.

\subsection{PCS-facing control state and process-loopback verification}

\begin{figure*}[t]
\centering
\includegraphics[width=0.96\textwidth]{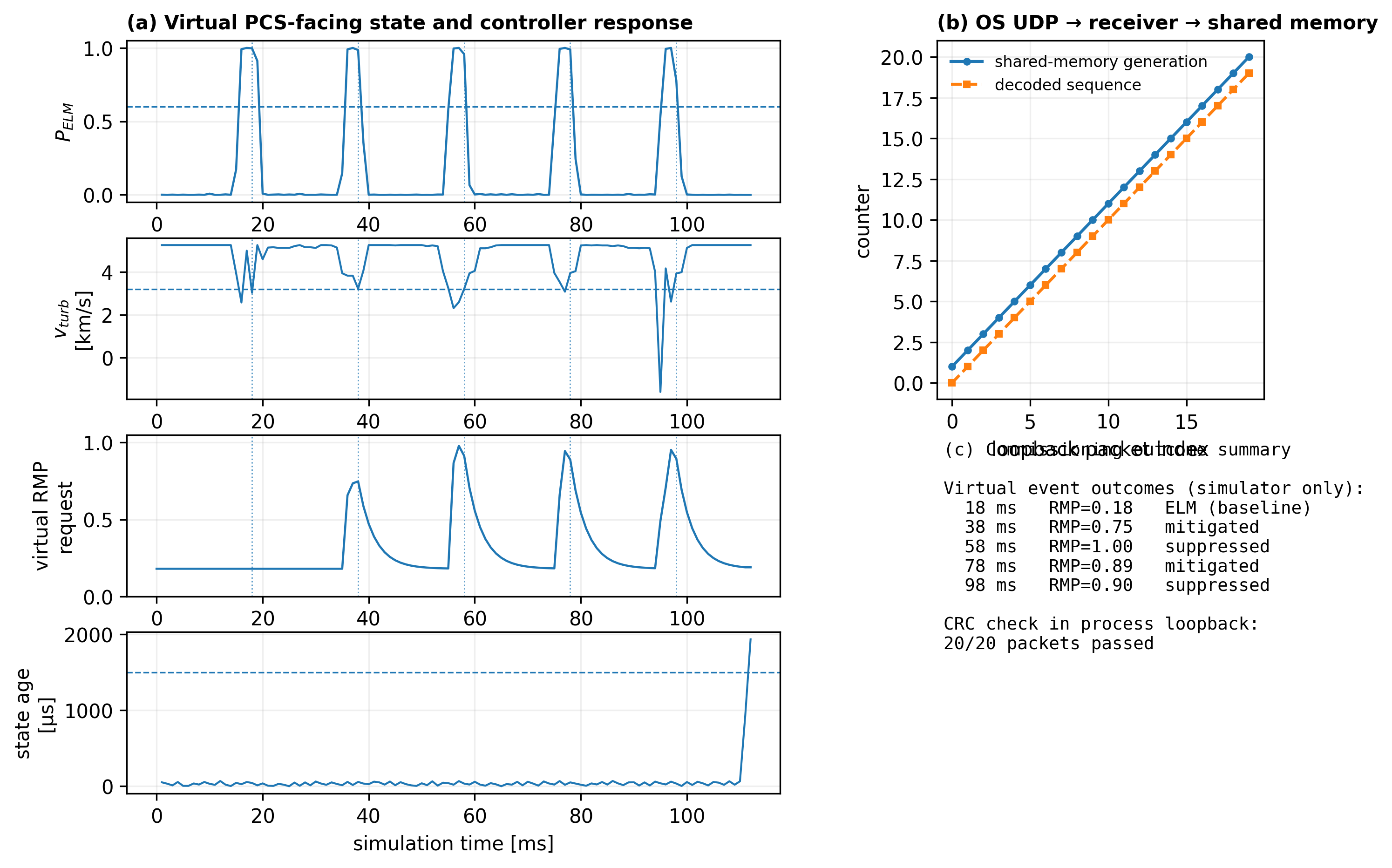}
\caption{PCS-facing simulator behavior and process-level loopback verification. (a) Synthetic edge outputs, virtual controller request, and shared-state age in the 1~kHz PCS-like loop. Event labels and actuator response are simulator-only. (b) Generation and sequence counters observed while binary result packets cross a localhost UDP receiver and POSIX shared-memory boundary. (c) Example virtual event outcomes and CRC summary. The final state-age increase occurs after the 110~ms synthetic data stream ends and demonstrates the stale-state logic.}
\label{fig:control}
\end{figure*}

Figure~\ref{fig:control} illustrates the downstream behavior generated from the same baseline run. The left panels show the synthetic $\pelm$, CCTD-like $\vturb$, virtual RMP request, and shared-state age sampled by the 1~kHz control task. The vertical event markers correspond to the synthetic candidate-ELM schedule. The reported ``mitigated'' and ``suppressed'' labels are produced by a deliberately heuristic virtual plant response and must not be interpreted as DIII-D control results. Their purpose is to force the data path to exercise different controller states and actuator levels.

The upper-right panel of Fig.~\ref{fig:control} shows a separate process-level test. Twenty binary FPGA-result packets are sent over a real localhost UDP socket to a receiver that writes POSIX shared memory. The reader observes monotonic sequence and generation counters, and CRC verification succeeds for 20/20 packets. This test does not establish deterministic real-time performance, but it verifies that the same binary packet can cross an operating-system socket boundary and populate a shared state with no special GUI or notebook dependency.

\section{Interactive commissioning interface}
\label{sec:gui}

A full-system simulator becomes more useful when a diagnostician, network engineer, FPGA developer, and control-system developer can change different parts of the same configuration without editing the underlying code. The primary GUI uses the same configuration and packet-processing functions.

The GUI exposes the diagnostic duration and seed; samples per packet and SPAD count; link jitter, random loss, duplication, and reordering; deterministic drop, delay, or counter-jump faults; FPGA filter, window, stride, and component delays; PCS network delay, control frequency, stale timeout, fail-safe mode, and control thresholds. Time-scrub controls show native $8\times8$ BES pixels as well as interpolated images and contours. Packet inspectors display HUDP metadata, SPAD0 values, ADC codes, the binary hex payload, result-packet fields, and CRC status. The same run can export HDF5 data, CSV traces, and PCAP traffic.

\begin{figure*}[t]
\centering
\includegraphics[width=0.94\textwidth]{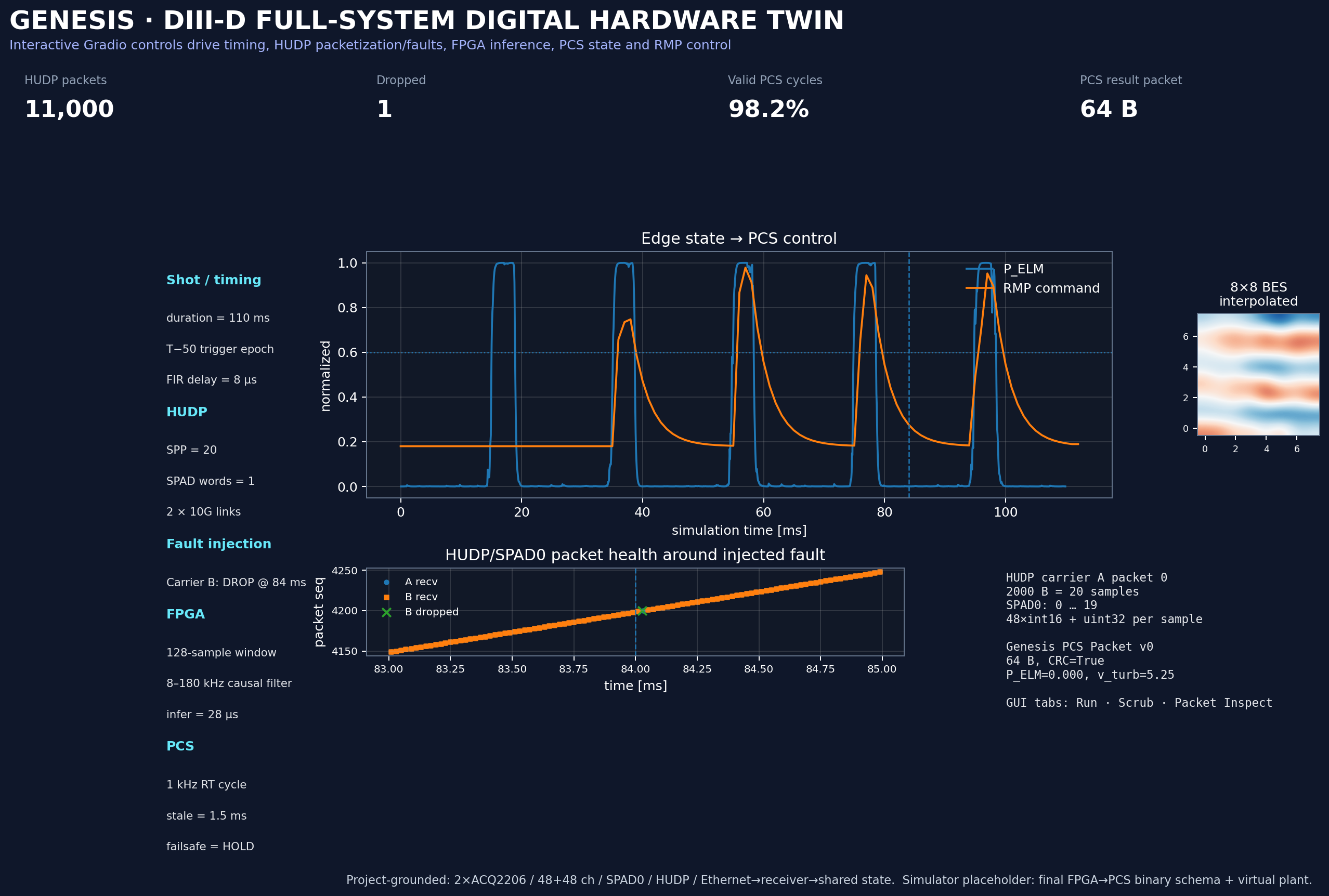}
\caption{Static view of the interactive commissioning interface generated from the same synthetic HDF5 and packet traces used in the paper. The operator can inspect $\pelm$, $\vturb$, the virtual actuator request, stream validity, latency statistics, and the spatial BES state. The current GUI is a development tool and not the production DIII-D PCS user interface.}
\label{fig:gui}
\end{figure*}

Figure~\ref{fig:gui} shows a static capture of the current operator view. The interface is not intended to mimic the production DIII-D PCS graphical environment. Its role is instead to provide an engineering commissioning surface in which packet health, inference state, timing, and controller behavior can be observed together before the final facility GUI is available.

\section{Discussion}
\label{sec:discussion}

\subsection{Relation to the deployed DIII-D framework}

The principal reason to target DIII-D in this first implementation is that the facility already provides a mature control system and an increasingly capable high-rate diagnostic framework. The present work addresses a different but complementary question: how can our acquisition/accelerator path be tested as an integrated system while the final transport firmware, receiver process, and facility deployment are still being completed?

The digital hardware twin is therefore most useful as an interface contract and regression environment. A change to samples-per-packet, counter semantics, packet byte order, filter group delay, model window, receiver timeout, or PCS packet version can be evaluated against the complete chain instead of against isolated unit tests. This is particularly relevant for FPGA-based ML, where a model that is numerically correct off-line can still fail to be useful in control if channels are reordered, timing epochs are inconsistent, packet faults are hidden, or a receiver publishes stale state.

\subsection{Why packet-level simulation is needed}

Array-level HIL demonstrations often assume that the accelerator receives an already synchronized tensor. That abstraction omits several failures that are likely during commissioning. For example, the two 48-channel carriers can have unequal packet delays; a packet can be dropped while subsequent packets continue; a counter can reset or jump; or a network receiver can deliver valid packets out of order. The ML input is only well defined after these effects have been resolved against the hardware time base. By representing the binary packet and counter explicitly, the present simulator can answer whether an input window is scientifically aligned before asking whether the neural-network prediction is correct.

This distinction is also important for future cross-machine deployment. A portable edge-AI model should not require the same digitizer or packet format at every tokamak. The appropriate abstraction is a device-specific acquisition adapter that produces a canonical synchronized diagnostic stream, followed by a reusable preprocessing/inference core. The digital hardware twin simulator already separates these modules, so a fusion reactor, laboratory replay, or other diagnostic transport can be added without modifying the control-facing state definition.

\subsection{Hardware measurements}

The present results have four important limitations. First, the latency distribution in Fig.~\ref{fig:latency} is a configured engineering model. It should be replaced component-by-component by FPGA timestamp measurements, network packet captures, receiver timestamps, and PCS shared-state observations. Second, the PCS Packet v0 is not the final DIII-D receiver schema. It is a versioned placeholder that permits software integration to proceed without claiming a facility contract that has not yet been finalized. Third, the synthetic BES source and logistic ELM surrogate are designed to exercise signal-processing and control interfaces; they do not establish predictive performance. Recorded DIII-D BES replay and the trained FPGA model are the next validation stages. Fourth, the virtual RMP/plant response is intentionally heuristic and is not a plasma-control model.

For those reasons, the planned validation sequence is staged. The first step is software-only regression with synthetic data, as reported here. The second is recorded-shot replay through the same packet and FPGA interfaces, with direct comparison against off-line BES preprocessing and model outputs. The third is a bench test with the physical D-TACQ and FPGA hardware, using common clock/trigger signals and packet captures to measure counter behavior, filter delay, and transport latency. The fourth is a PCS receiver/shared-memory test on the DIII-D network without actuator authority. Only after these stages should the path be enabled in a control experiment. This staged approach is consistent with the broader principle of separating high-bandwidth experimental processing from the deterministic safety-critical PCS core \cite{kim2026rtmonitoring,abbate2023infrastructure}.

The simulator is primarily an engineering design and integration tool. Its central objects are diagnostic acquisition, timing contracts, network packetization, accelerator interfaces, fault handling, shared-memory publication, and control-system commissioning. The software also provides a reproducible method for comparing implementation alternatives before procurement or machine time is committed, including alternative packet sizes, accelerator latencies, network paths, and fail-safe policies.

\section{Conclusion}
\label{sec:conclusion}

A DIII-D-oriented, packet-level digital hardware twin has been developed for commissioning the megahertz diagnostic edge-AI path from megahertz BES acquisition to a PCS-facing control state. The simulator embeds 64 active BES channels in a 96-channel dual-carrier system, generates binary HUDP packets with SPAD0 sample counters, models independent link timing and faults, reconstructs a synchronized FPGA input stream, applies causal preprocessing and edge inference, encodes a versioned result packet, writes receiver shared state, and exercises a 1~kHz virtual control cycle.

In the baseline 110~ms test, the two carriers generated 11,000 packets. A single intentionally dropped 20-sample carrier packet was detected through counter continuity and invalidated the two overlapping 128-sample inference windows. For valid windows, the configured last-input-to-shared-memory latency had a median of 91.6~$\mu$s and a 99th percentile of 108.1~$\mu$s. A separate UDP-to-POSIX-shared-memory process test preserved sequence and CRC for all 20 transmitted result packets. These results verify software continuity and fault propagation.

The main value of the framework is that system integration can begin while the final FPGA bitstream, receiver process, and PCS schema are still being finalized. The next steps are to replace the synthetic source with recorded DIII-D BES replay, ingest packet captures from the physical D-TACQ hardware, and substitute the final DIII-D receiver/shared-memory contract. With these replacements, the same software structure can evolve from an expected-system simulator into a deployment-level commissioning twin and can subsequently be adapted to other fusion devices through acquisition-specific interface modules.

\section*{Declaration of competing interest}
The authors declare no competing interests.

\section*{Acknowledgments}
This work is supported by the U.S. Department of Energy, Office of Science, Office of Fusion Energy Sciences, using the DIII-D National Fusion Facility, a DOE Office of Science user facility, under Award(s) DE-FC02-04ER54698, DE-SC0021157, DE-SC0001288, DE-SC0020287, and DE-FG02-08ER54999. This research was supported by the Scientific Discovery through Advanced Computing (SciDAC) program, specifically Center for Edge of Tokamak OPtimization (CETOP), under Award Number DE-AC02-09CH11466. This research used resources of the National Energy Research Scientific Computing Center (NERSC), a U.S. Department of Energy Office of Science User Facility located at Lawrence Berkeley National Laboratory, operated under Contract No. DE-AC02-05CH11231 using NERSC Award FES-ERCAP0026282. Supported by Genesis awards from the US Department of Energy, Office of Science, though this abstract was prepared prior to the issuance of award numbers.

Disclaimer: This report was prepared as an account of work sponsored by an agency of the United States Government. Neither the United States Government nor any agency thereof, nor any of their employees, makes any warranty, express or implied, or assumes any legal liability or responsibility for the accuracy, completeness, or usefulness of any information, apparatus, product, or process disclosed, or represents that its use would not infringe privately owned rights. Reference herein to any specific commercial product, process, or service by trade name, trademark, manufacturer, or otherwise does not necessarily constitute or imply its endorsement, recommendation, or favoring by the United States Government or any agency thereof. The views and opinions of authors expressed herein do not necessarily state or reflect those of the United States Government or any agency thereof.

\section*{Data and code availability}
The data that support the findings of this study are available from the corresponding author upon request. A public repository and archival release will be assigned after journal publication.

\end{document}